\documentclass[a4paper,12pt]{article}
\usepackage{soul}
\usepackage[usenames,dvipsnames]{color}
\usepackage{jheppub}
\usepackage{amsmath, amssymb, slashed, epsf, color, graphicx, latexsym}
\usepackage{epsfig}
\usepackage{graphics}
\newcommand{\CP}{{\cal P}}

\newcommand{\mK}{\mathcal{K}}

\begin{document}
	
	\preprint{TIFR/TH/16-36}
	\title{Unstable `black branes' from scaled membranes at large $D$}
	\author[a]{Yogesh Dandekar,} 
	\author[a]{Subhajit Mazumdar,}
	\author[a]{Shiraz Minwalla}
	\author[a]{and Arunabha Saha}
	\affiliation[a]{Tata Institute of Fundamental Research, Mumbai, India-400005}
	\emailAdd{yogesh@theory.tifr.res.in}
	\emailAdd{subhajitmazumdar@theory.tifr.res.in}
	\emailAdd{minwalla@theory.tifr.res.in}
	\emailAdd{arunabha@theory.tifr.res.in}
	
	\abstract{It has recently been demonstrated that the dynamics of 
black holes at large $D$ can be 
recast as a set of non gravitational membrane 
equations. These membrane equations admit a simple static 
solution with shape $S^{D-p-2} \times R^{p,1}$. In this note we study the 
equations for small fluctuations about this solution in a 
limit in which amplitude and length scale of the  fluctuations are 
simultaneously scaled to zero as $D$ is taken to infinity. 
We demonstrate that the resultant nonlinear equations, which capture 
the Gregory- Laflamme instability and its end point, exactly agree with the effective 
dynamical `black brane' equations of Emparan Suzuki and Tanabe. Our results thus identify the 
`black brane' equations as a special limit of the membrane equations and so unify
these approaches to large $D$ black hole dynamics. }

%opening
\maketitle

%\tableofcontents
\section{Introduction}

A few years ago Emparan, Suzuki, Tanabe and collaborators 
observed \cite{Emparan:2013moa,Emparan:2013xia,Emparan:2013oza,Emparan:2014cia,Emparan:2014jca,Emparan:2014aba,Emparan:2015rva}(see also\cite{Giribet:2013wia,Prester:2013gxa}) that the 
classical equations that govern the dynamics of black holes in $D$ 
dimensions simplify in the large $D$ limit. Motivated by this observation, 
several papers written over the last year or so have demonstrated that 
black hole physics at large $D$ can be reformulated in terms of  dual non 
gravitational equations. In broad terms there have been two different 
approaches to this problem.

The first of these approaches is laid out in the `membrane paradigm' 
papers of \cite{Bhattacharyya:2015dva, Bhattacharyya:2015fdk, Dandekar:2016fvw} (see also \cite{Emparan:2015hwa, Suzuki:2015iha, Tanabe:2015hda} \footnote{
These papers worked out the equations that govern the shape of the membrane, 
described later in this paragraph, for stationary configurations. Atleast 
in the absence of a cosmological constant, these  equations may be shown 
to follow from the more general dynamical membrane equations of 
\cite{Bhattacharyya:2015dva, Bhattacharyya:2015fdk, Dandekar:2016fvw}
upon inserting an appropriate stationary ansatz, and so are special cases 
of the general membrane equations.}) . 
The authors of these papers have demonstrated  
that nonlinear black hole dynamics can be reformulated in terms of 
the equations of motion of a non gravitational membrane that lives 
in flat space. The variables of this problem are the shape of the 
membrane and a velocity field on this membrane\footnote{The variables of 
the membrane also include a charge field for charged black holes. In this 
note, however, we focus solely on solutions of the vacuum Einstein equations 
$R_{MN}=0$. We leave the generalization of our study to charged black holes,  
and to dynamics in spaces with nonzero cosmological constants,  to future 
work.}. Einstein's equations force the 
membrane variables to obey a set of equations of motion. There are
as many equations of motion as variables, so the membrane description 
defines a good initial value problem. We emphasize that 
the membrane equations of \cite{Bhattacharyya:2015dva, Bhattacharyya:2015fdk, Dandekar:2016fvw} apply to arbitrarily nonlinear and completely 
dynamical black hole motions. There are, in particular, no restrictions 
on the initial shape of the membrane which can be chosen to be any sufficiently 
smooth codimension one submanifold of  flat spacetime; the evolution 
of this shape (and the membrane velocity fields) in time is, of course, 
governed by the membrane equations of motion. 

A second approach is that of the `scaled black brane' papers of 
\cite{Emparan:2015gva,Emparan:2016sjk} (see also \cite{Suzuki:2015axa,
Tanabe:2016pjr}). These papers  study small fluctuations 
about the $p$ dimensional `black brane'; a spacetime given by the 
direct product of the Schwarzschild solution in $R^{D-p-1, 1}$ 
and $R^p$. The authors of \cite{Emparan:2015gva} consider fluctuations that 
preserve $SO(D-p-1)$ isometry but vary in the $R^p$ direction over length scales of 
order $\frac{1}{\sqrt{D}}$ and time scales of order unity. Focusing 
attention on wiggles of the event horizon of amplitude $\frac{1}{D}$ and 
on boost velocities of the horizon of order $\frac{1}{\sqrt{D}}$, the authors
of  \cite{Emparan:2015gva} were able to derive a set of effective non 
gravitational 
nonlinear equations that completely reproduce black brane dynamics in 
the scaled large $D$ limit described above. This scaling limit is of 
particular interest because it turns out to capture the Gregory-Laflamme 
instability of black branes at large $D$.  \footnote{  
\cite{Emparan:2015gva} has subsequently been generalized to 
the the study of charged black branes in 
\cite{Emparan:2016sjk}. As mentioned above, however, in this note we 
focus attention on uncharged black holes and black branes.}
\footnote{Additional recent studies of black hole physics at large $D$ 
include 
\cite{Tanabe:2016opw,Sadhu:2016ynd,Herzog:2016hob,Rozali:2016yhw,Chen:2015fuf,Chen:2016fuy}).}

In this brief note we derive the `black brane' equations of 
\cite{Emparan:2015gva} starting from the membrane equations of 
\cite{Bhattacharyya:2015dva, Bhattacharyya:2015fdk, Dandekar:2016fvw}.
The starting point of our analysis is the simple exact solution 
to the membrane equations of motion that is  dual to the $p$ dimensional 
`black brane' described in the previous paragraph. This solution is 
static, which means that the membrane 
velocity field is simply given by $u=-dt$. The shape of the membrane
on this solution is $S^{D-p-2} \times R^{p,1}$. We then proceed 
to study the scaling limit of \cite{Emparan:2015gva} directly within
the membrane picture. In other words we study fluctuations of the membrane
that preserve $SO(D-p-1)$ isometry but vary in the $R^p$ direction over 
length scales of order $\frac{1}{\sqrt{D}}$ and time scales of order unity. 
We then focus on wiggles of the shape of the membrane with amplitude 
of order 
$\frac{1}{D}$ and on membrane velocities of order $\frac{1}{\sqrt{D}}$. 
At leading order in the large $D$ limit we obtain a simple set of scaled 
equations of membrane dynamics which (after the appropriate field 
redefinitions) turn out to agree exactly with the equations of 
 \cite{Emparan:2015gva}. We view our derivation of the (uncharged) black 
brane equations from the membrane equations as a unification of these 
two approaches to horizon dynamics at large $D$. Note it follows, in 
particular, that the dynamics of the Gregory Laflamme instability
is captured by scaling limit of membrane equations described above.

The limit of the previous paragraph is loosely
reminiscent  of the scaling limit that yields the nonrelativistic 
Navier-Stokes equations starting from the more general relativistic equations \cite{Bhattacharyya:2008kq}. The membrane equations may also admit other 
interesting scaling limits. We leave the investigation of this point to 
future work. 

\section{A scaling limit of the membrane equations}

 In this note we study the  equations of 
motion \cite{Bhattacharyya:2015dva,Bhattacharyya:2015fdk,Dandekar:2016fvw} of an uncharged large D membrane propagating in flat Minkowski spacetime. 
To leading order in $\frac{1}{D}$ these equations take the form 
\begin{equation} \label{veq1geomwv}
\Bigg[\frac{\nabla^{2}u_{A}}{\mathcal{K}}-\frac{\nabla_{A} \mathcal{K}}{\mathcal{K}}+u^{B} K_{B A}-u^{B}\nabla_{B} u_{A}\Bigg]\CP^{A}_{C}=0 
\end{equation}
with ,
\begin{equation}\label{divel}
\nabla.u=0.
\end{equation}
Here $K_{AB}$ is the extrinsic curvature of the membrane, $\mK$ is its trace and $u$ is the local world volume velocity field of the membrane. All covariant derivatives in \eqref{veq1geomwv} and \eqref{divel} are defined 
with respect to the induced metric on the membrane. Also  
\begin{equation}
\CP^{A B}= \hat{g}^{A B} + u^{A} u^{B} 
\end{equation}
where $\hat{g}^{AB}$ is the metric induced from the ambient flat space 
on the world volume of the membrane. 
In other words $\CP^{AB}$ is the projector, on the membrane world volume, 
orthogonal to the velocity field $u$.

\subsection{Linearized Fluctuations}

In our study we will find it useful to use coordinates in which the flat 
space $D$ dimensional metric takes the form 
\begin{equation}\label{metric}
 ds^2=-dt^2+d\tilde{x}^a d\tilde{x}^a+dr^2+r^2d\Omega^2_{n}
\end{equation}
where
$$n=D-p-2.$$
and $a=1\ldots p$ label the spatial directions on the black brane. 
A simple solution to the equations \eqref{veq1geomwv} 
and \eqref{divel} 
is given by the membrane shape $r=1$ and constant static velocity 
field $u=-dt$. \footnote{The choice $r=1$ involves 
no loss of generality, as the scale invariance 
of the classical Einstein equations relate the solution with 
$r=1$ to the solution with $r=r_0$ for any constant $r_0$.}

The solution of the membrane equations described in the previous 
paragraph is dual to a `black brane' - the solution of general 
relativity given by the direct product of $R^p$ and 
the Schwarschild black hole in $D-p$ dimensions. It is 
well known that this solution of general relativity is unstable 
in an arbitrary number of dimensions. We will now use the membrane 
equations to exhibit this instability, by linearizing these equations
about the simple solution. The Gregorry Laflamme instability of black 
branes is known to preserve the $SO(n+1)$ symmetry of the sphere but
to break translational invariance along $R^p$, so we study 
fluctuations with the same property. In other words we set 
\begin{equation}
\begin{split}
 &r=1+ \tilde{\delta r}(t,\tilde{x}^a) \\
 &u=-dt+ \tilde{\delta u_a}(t,\tilde{x}^a)d\tilde{x}^a 
\end{split}
\end{equation}
Note that our velocity fluctuations lie entirely in the black brane directions
and none of our fluctuations fields depend on the angular variables 
on $S^n$. 

Following the method described in section 5 of \cite{Bhattacharyya:2015fdk}, 
it is not difficult to linearize the membrane equations around the `black brane'
solution.  The 
equation \eqref{divel} reduces to 
\begin{equation}\label{laphi}
 n~\partial_t \tilde{\delta r} + \tilde{\partial}_a \tilde{\delta u^a}=0
\end{equation}
(recall $n=D-p-2$). 
\footnote{The factor of $n$, which plays a crucial role in the 
analysis below, has its origin in the fact that the induced metric on the 
world volume of the membrane is given, to leading order in fluctuations by  
$$ ds^2= -dt^2 + d {\tilde x}^a d {\tilde x}_a+ (1+ 2 \tilde{\delta r}) d \Omega_{n}^2 $$
so that $\sqrt{g}=1+ n \tilde{\delta r}$ in these coordinates. } 
The equation with a free index in the (spatial) $R^p$ 
direction turns out to take the form
\begin{equation}\label{zeq}
 \begin{split}
  & \left( \tilde{\partial}_a {\tilde \delta r } 
  - \partial_t\tilde{\partial}_a \tilde{\delta r} - \partial_t \tilde{\delta u_a} \right)  + \left( \frac{-\partial^2_t+\tilde{\partial}_b\tilde{\partial}^b}{n} 
\right) \left( \tilde{\delta u_a} + {\tilde \partial}_a 
{\tilde \delta r} \right)  =0
 \end{split}
\end{equation}
where $\tilde{\partial}_a$ is the derivative with respect to the coordinate $\tilde{x}^a$ defined in \eqref{metric}.
When all spatial and time derivatives are of order unity or smaller, the term 
$$\left( \frac{-\partial^2_t+\tilde{\partial}_b\tilde{\partial}^b}{n} 
\right) \left( \tilde{\delta u_a} + {\tilde \partial}_a 
{\tilde \delta r} \right)$$
in \eqref{zeq} is subleading in the $\frac{1}{n}$ expansion and so 
can naively be dropped at leading order.  
However we will soon find ourselves interested in configurations with 
spatial derivatives of order $\sqrt{n}$ but time derivatives of order unity. 
For such configurations the term proportional to time derivatives in 
\eqref{zeq} is indeed subleading in $\frac{1}{n}$. On the other hand 
the term proportional to 
the spatial laplacian is comparable to the other terms in \eqref{zeq} 
and so must be retained. Over the parameter ranges of interest to this paper, 
therefore, we can replace \eqref{zeq} with the slightly simpler equation
\begin{equation}\label{zeqm}
 \begin{split}
 \left( \tilde{\partial}_a {\tilde \delta r } 
  - \partial_t\tilde{\partial}_a \tilde{\delta r} - \partial_t \tilde{\delta u_a} \right)  + \left( \frac{\tilde{\partial}_b\tilde{\partial}^b}{n} 
\right) \left( \tilde{\delta u_a} + {\tilde \partial}_a 
{\tilde \delta r} \right)  =0
 \end{split}
\end{equation}
\footnote{The membrane equations with free index in sphere direction is trivially 
satisfied, while the equation in the time direction is also a triviality 
(this is a consequence of the projector in \eqref{veq1geomwv}).}

The equations \eqref{zeqm} and \eqref{laphi} are easily analysed. 
Substituting the plane wave expansion 
\begin{eqnarray}\label{mode_exp}
&&\tilde{\delta r}(t,\tilde{x}^a)= \delta r^0 e^{-i \omega t}e^{i\tilde{k}_a\tilde{x}^a}\nonumber\\
&&\tilde{\delta u_a}(t,\tilde{x}^a) = \delta u_a^0 e^{-i \omega t}e^{i\tilde{k}_a\tilde{x}^a}
\end{eqnarray} 
into \eqref{zeqm} and \eqref{laphi} turns these equations into eigenvalue 
equations for the fluctuation frequencies $\omega$. Solving the resultant 
cubic equation in $\omega$ we find find that the most general 
solution to these equations is given by 
\begin{equation}\begin{split}\label{linsol}
& \tilde{\delta r}(t,\tilde{x}^a) = \delta r_1^0 e^{-i \omega_1 t}e^{i\tilde{k}_a\tilde{x}^a} + \delta r_2^0 e^{-i \omega_2 t}e^{i\tilde{k}_a\tilde{x}^a}\\
& \tilde{\delta u_a}(t,\tilde{x}^a) = \delta r_1^0 \tilde{k}_a \left( - i + \frac{\sqrt{n}}{\tilde{k}} \right)  e^{-i \omega_1 t}e^{i\tilde{k}_a\tilde{x}^a}
+  \delta r_2^0 \tilde{k}_a \left( - i - \frac{\sqrt{n}}{\tilde{k}} \right)  e^{-i \omega_2 t}e^{i\tilde{k}_a\tilde{x}^a}
+ v_a e^{- i \omega_3 t}e^{i\tilde{k}_a\tilde{x}^a}\\
& w_1=i\left(\frac{\tilde{k}}{\sqrt{n}} - \frac{\tilde{k}^2}{n}\right), ~~~
w_2=i\left(-\frac{\tilde{k}}{\sqrt{n}} - \frac{\tilde{k}^2}{n}\right), ~~~
w_3=-i\frac{\tilde{k}^2}{n}, ~~~{\rm where}~~~ {\tilde k}^2= {\tilde k}_a 
{\tilde k}^a
\end{split}
\end{equation}
\eqref{linsol} is a solution to the linearized membrane equations for 
arbitrary constant values of $\delta r_1^0$ and $\delta r_2^0$  and for any 
constant vector $v_a$ s.t. $\tilde{k}^a v_a=0$. 

Note that the mode proportional to $\delta r_1^0$ - i.e. the mode with frequency 
$\omega_1$ - is unstable when ${\tilde k}<\sqrt{n}$. This IR instability (i.e. an instability 
that occurs at distance scales lareger than a minimum) is the membrane dual of the Gregory Laflamme instability. When  ${\tilde k}$ is of order unity time scale associated with this frequency is of order $\sqrt{n}$ and so is very large. The minimum time scale for an instability, however, occurs at 
${\tilde k}= \frac{\sqrt{n}}{2}$. At this wavelength the time scale of the instability is 
order unity. \footnote{   The expression for the unstable mode $w_1$ was conjectured earlier from fluid/gravity methods in \cite{Camps:2010br}. See also \cite{Caldarelli:2012hy} and \cite{Emparan:2013moa} for further evidence for the above proposal. 
}

At the level of the linearized equations the Gregory - Laflamme unstable modes simply grow forever. 
Nonlinear effects, however, stabilize these modes. The discussion of the previous paragraph makes 
it clear that the length scale relevant to this physics is $\frac{1}{\sqrt{n}}$. We will now 
proceed to find the effective nonlinear theory within which the Gregory- Laflamme instability and 
its end point can be reliably studied.

\subsection{Scaled nonlinear equations}

In order to restrict attention to distance of order $\frac{1}{\sqrt{n}}$ in the 
spatial black brane directions we work with the scaled coordinate
$x^a$ defined by $\tilde{x}^a =  \frac{x^a}{\sqrt{n}}$. 
Unstable modes with finite wavelength in this new coordinate have 
frequencies of order unity. The background flat space metric now takes the form 
\begin{equation}\label{smo} 
ds^2=-dt^2+dr^2+ \frac{1}{n} dx_{a}dx^{a}+r^2 d \Omega_{n}^2
\end{equation}
As our fluctuations field all vary over distances of order unity and 
time scales of order unity in scaled coordinates, the velocity  field 
$u^a$ should thus also be of order unity. This implies that $u_a \sim {\cal O}
(\frac{1}{n}$) . Translating back to unscaled coordinates it follows that 
${\tilde u}^a= {\cal O}(\frac{1}{\sqrt{n}})$. In order to ensure this 
scaling in our solution \eqref{linsol} we must choose
 $v_a \sim {\cal O}(\frac{1}{\sqrt{n}})$,  
$\delta r_1^0 \sim \delta r_2^0 \sim {\cal O}(\frac{1}{n} )$. 
These choices, in turn, ensure that 
$\tilde{\delta r} \sim {\cal O}(\frac{1}{n})$
(see \eqref{linsol}). It is thus natural to make the further coordinate change 
\begin{equation}\label{cc}
r= 1+ \frac{y}{n}
\end{equation}
The flat space metric is now given by 
\begin{equation}\label{sm} 
ds^2=-dt^2+\frac{dy^2}{n^2}+ \frac{1}{n} dx_{a}dx^{a}+ 
\left( 1 + \frac{y}{n} \right)^2 d \Omega_{n}^2
\end{equation}

With our scalings now in place we focus attention on membrane configurations of the form 
\begin{equation}\label{sconfig} \begin{split}
y&=y(x^a, t)\\
u^a&=u^a(x^a, t)\\
\end{split}
\end{equation}
where the functions $y(x^a, t)$ and $u^a(x^a, t)$ are independent of $n$. We then evaluate 
the membrane equations \eqref{divel} and \eqref{veq1geomwv} for such configurations propagating on the metric 
\eqref{sm}. Retaining only terms of leading order at large $n$ we find that the equation 
\eqref{divel} (which we call $E^s$ below) and the $a$ components of \eqref{veq1geomwv} 
(which we call $E^v_a$ below) reduce to 
\begin{equation}\label{oureqn}
\begin{split}
 E^s&\equiv u^{b}\partial_{b}y+\partial^{b}u_{b}+\partial_{t}y=0\\
 E^v_a&\equiv \partial^b \partial_{b}u_{a}+\partial_{a}y-u^{b}\partial_{b} u_{a}+\partial^{b}y \partial_{b}u_{a}-u^{b}\partial_{b}\partial_{a}y\\ &+\partial^{b}y \partial_{b}\partial_{a}y+\partial^{b}\partial_{b}\partial_{a}y-\partial_{t}u_{a}-\partial_{t}\partial_{a}y=0
 \end{split}
\end{equation}
Note that the equations \eqref{oureqn} are nonlinear. If we linearize these equations around 
the background $y=u^a=0$ we obtain the linearized equations 
\begin{equation}\label{oureqnlin}
\begin{split}
  &\partial^{b}\delta u_{b}+\partial_{t}\delta r=0\\
 &\partial^b \partial_{b}\delta u_{a}+\partial_{a}\delta r+\partial^{b}\partial_{b}\partial_{a}\delta r-\partial_{t}\delta u_{a}-\partial_{t}\partial_{a}\delta r=0
 \end{split}
\end{equation}
The first and second of \eqref{oureqnlin} are simply \eqref{laphi} and \eqref{zeqm} expressed 
in scaled variables. It follows that \eqref{oureqn} are nonlinear 
generalizations of the linearized fluctuation equations of the 
previous subsection. The \eqref{oureqnlin} are exact at large $n$ 
within the scaling limit described in this section. 

The nonlinear equations \eqref{oureqn} capture both the linear exponential 
growth as well as the nonlinear settling down of the Gregory Laflamme instability. 
We do not need to perform the analysis of this fact, however, because it has 
already been done! We will now demonstrate that the equations \eqref{oureqn} are 
equivalent to those that Emparan Suzuki and Tanabe \cite{Emparan:2015gva} derived 
to study large $D$ `black branes' - and used to perform an extensive study of the 
Gregory Laflamme instability.

In order to make contact with the work of \cite{Emparan:2015gva} we make the following field redefinitions
\begin{eqnarray}
&&y(t,x^a)=\log m(t,x^a)\nonumber\\
&&u_a(t,x^a)=\frac{p_a(t,x^a)-\partial_a\left(m(t,x^a)\right)}{m(t,x^a)}
\end{eqnarray}
and work with the following linear combinations of \eqref{oureqn} 
\begin{equation}
E_1=m(t,x^a)E^s \quad \text{and} \quad E_a=p_a(t,x^a)  E^s- m(t,x^a) E^v_a.
\end{equation}
It is easily verified that $E_1$ and $E_a$ take the form
\begin{equation}\label{emeqn}
\begin{split}
&E_1=\partial_{t}m-\partial_{b}\partial^{b}m+\partial_{b}p^{b}=0\\
&E_a=\partial_{t}p_{a}-\partial_{b}\partial^{b}p_{a}-\partial_{a}m+
\partial_{b}\left(\frac{p_{a}p^{b}}{m}\right)=0
\end{split}
\end{equation}
The equations \eqref{emeqn} are precisely the nonlinear black brane equations $(11)$ and $(12)$ 
of \cite{Emparan:2015gva}. It follows that these black brane equations are simply a particular 
scaled limit of the general leading order (in an expansion in $\frac{1}{D}$) equations 
\eqref{divel} and \eqref{veq1geomwv}.

\section{Discussion}

In this note we have demonstrated by explicit computation that the uncharged `black brane' 
equations of \cite{Emparan:2015gva} may be obtained from a scaling limit of the general 
membrane equations \eqref{divel} and \eqref{veq1geomwv}. The reader may, at first, 
find herself puzzled at this 
agreement, given the scaling limit described in this note focuses on length scales of order 
$\frac{1}{\sqrt{D}}$ while that the membrane equations \eqref{divel} and \eqref{veq1geomwv} were derived as the first 
term in a systematic expansion in $\frac{1}{D}$ under the assumption that the horizon and 
velocity fields all vary on length scale unity. We will now explain why this agreement 
was infact to be expected despite the apparent conflict of regimes of validity. 

The equations \eqref{divel} and \eqref{veq1geomwv} would fail to accurately capture dynamics at leading order in the 
large $D$ limit if the explicit factors of $D$ in the metric \eqref{sm} ensured that a
higher order term\footnote{i.e. a term that appears at higher order in the expansion in $\frac{1}{D}$ in the membrane equations of \cite{Dandekar:2016fvw}}
were to contribute to the equations at same (or higher) order in $\frac{1}{D}$ as the terms in 
\eqref{divel} and \eqref{veq1geomwv}. We will now explain that this never happens. Potentially dangerous terms are 
those that contain one or more factors of the inverse metric $g^{ab}$ where 
the indices $a$ and $b$ are spatial black brane directions. These terms are 
potentially dangerous as $g^{ab}$ (see \eqref{sm}) is of order $D$. 
However these factors never actually lead to a mixing of orders 
because the extra indices $a$ and $b$ each need to contract with something. 
When these indices contract with $u_a$ the extra factor of $D$ is nullified 
by the fact that $u_a$ is of order $\frac{1}{D}$. When these indices 
contract with a derivative, the derivative acts on some quantity built 
out of fluctuation fields. However all such quantities are of order 
$\frac{1}{D}$ (recall, for instance, that every fluctuation component of 
the extrinsic curvature is proportional to ${\tilde \delta r}$ which is 
of order $\frac{1}{D}$). The smallness of fluctuations in our scaling limit 
once again counteracts the potential enhancement of powers of $D$. 
It follows that leading order equations \eqref{divel} and \eqref{veq1geomwv} is infact 
sufficient to capture the leading order large $D$ dynamics of the scaling 
limit described in this note despite the fact that the scaling limit zooms 
in on distance scales of order $\frac{1}{\sqrt{D}}$. 

It should not be difficult to generalize the discussion of this note 
to obtain the first corrections,  in an expansion in $\frac{1}{D}$, 
to the black brane equations of \eqref{emeqn}.  These corrections have been obtained from `scaled black brane' approach in \cite{Herzog:2016hob,Suzuki:2015axa,Rozali:2016yhw}.  The starting point for such an analysis would be 
the first order corrected membrane equations derived in \cite{Dandekar:2016fvw}. 
It would also be interesting to check whether the analysis of this note generalizes to a derivation 
of the charged `black brane' equations of \cite{Emparan:2016sjk} starting with the charged membrane equations 
of \cite{Bhattacharyya:2015fdk}. \footnote{The discussion of the last 
paragraph suggests that this is guaranteed to work only if the scaling limit of 
\cite{Emparan:2016sjk} turns on membrane charge fluctuations that scale 
like $\frac{1}{D}$. }
We leave a study of these issues to future work.

We end this note by reiterating that we have demonstrated that the 
black brane equations of \cite{Emparan:2015gva} can be derived as a special case of the more general membrane 
equations of \cite{Bhattacharyya:2015dva,Bhattacharyya:2015fdk,Dandekar:2016fvw}, leading to a satisfying unification recent attempts to 
reformulate large $D$ horizon dynamics in non gravitational terms.

\section*{Acknowledgments}
We would like  to 
thank S. Bhattacharyya for several very useful discussions and explanations. 
The work
of all authors was supported by the Infosys Endowment for the study of the 
Quantum Structure of Spacetime, as well as an Indo Israel (UGC/ISF) grant. 
Finally we would all like to acknowledge our debt to the people of India 
for their steady and generous support to research in the basic sciences.

%--------------------------------------------
%--------------------------------------------

%--------------------------------------------
%-------------------------------------------

\bibliographystyle{JHEP}
\bibliography{ssbib}

\providecommand{\href}[2]{#2}\begingroup\raggedright\begin{thebibliography}{10}

\bibitem{Emparan:2013moa}
R.~Emparan, R.~Suzuki and K.~Tanabe, \emph{{The large D limit of General
  Relativity}}, \href{http://dx.doi.org/10.1007/JHEP06(2013)009}{\emph{JHEP}
  {\bf 1306} (2013) 009}, [\href{http://arxiv.org/abs/1302.6382}{{\tt
  1302.6382}}].

\bibitem{Emparan:2013xia}
R.~Emparan, D.~Grumiller and K.~Tanabe, \emph{{Large-D gravity and low-D
  strings}},
  \href{http://dx.doi.org/10.1103/PhysRevLett.110.251102}{\emph{Phys.Rev.Lett.}
  {\bf 110} (2013) 251102}, [\href{http://arxiv.org/abs/1303.1995}{{\tt
  1303.1995}}].

\bibitem{Emparan:2013oza}
R.~Emparan and K.~Tanabe, \emph{{Holographic superconductivity in the large D
  expansion}}, \href{http://dx.doi.org/10.1007/JHEP01(2014)145}{\emph{JHEP}
  {\bf 1401} (2014) 145}, [\href{http://arxiv.org/abs/1312.1108}{{\tt
  1312.1108}}].

\bibitem{Emparan:2014cia}
R.~Emparan and K.~Tanabe, \emph{{Universal quasinormal modes of large D black
  holes}}, \href{http://dx.doi.org/10.1103/PhysRevD.89.064028}{\emph{Phys.Rev.}
  {\bf D89} (2014) 064028}, [\href{http://arxiv.org/abs/1401.1957}{{\tt
  1401.1957}}].

\bibitem{Emparan:2014jca}
R.~Emparan, R.~Suzuki and K.~Tanabe, \emph{{Instability of rotating black
  holes: large D analysis}},
  \href{http://dx.doi.org/10.1007/JHEP06(2014)106}{\emph{JHEP} {\bf 1406}
  (2014) 106}, [\href{http://arxiv.org/abs/1402.6215}{{\tt 1402.6215}}].

\bibitem{Emparan:2014aba}
R.~Emparan, R.~Suzuki and K.~Tanabe, \emph{{Decoupling and non-decoupling
  dynamics of large $D$ black holes}},
  \href{http://dx.doi.org/10.1007/JHEP07(2014)113}{\emph{JHEP} {\bf 1407}
  (2014) 113}, [\href{http://arxiv.org/abs/1406.1258}{{\tt 1406.1258}}].

\bibitem{Emparan:2015rva}
R.~Emparan, R.~Suzuki and K.~Tanabe, \emph{{Quasinormal modes of (Anti-)de
  Sitter black holes in the 1/D expansion}},
  \href{http://arxiv.org/abs/1502.02820}{{\tt 1502.02820}}.

\bibitem{Giribet:2013wia}
G.~Giribet, \emph{{Large D limit of dimensionally continued gravity}},
  \href{http://dx.doi.org/10.1103/PhysRevD.87.107504}{\emph{Phys. Rev.} {\bf
  D87} (2013) 107504}, [\href{http://arxiv.org/abs/1303.1982}{{\tt
  1303.1982}}].

\bibitem{Prester:2013gxa}
P.~D. Prester, \emph{{Small black holes in the large D limit}},
  \href{http://dx.doi.org/10.1007/JHEP06(2013)070}{\emph{JHEP} {\bf 06} (2013)
  070}, [\href{http://arxiv.org/abs/1304.7288}{{\tt 1304.7288}}].

\bibitem{Bhattacharyya:2015dva}
S.~Bhattacharyya, A.~De, S.~Minwalla, R.~Mohan and A.~Saha, \emph{{A membrane
  paradigm at large D}},
  \href{http://dx.doi.org/10.1007/JHEP04(2016)076}{\emph{JHEP} {\bf 04} (2016)
  076}, [\href{http://arxiv.org/abs/1504.06613}{{\tt 1504.06613}}].

\bibitem{Bhattacharyya:2015fdk}
S.~Bhattacharyya, M.~Mandlik, S.~Minwalla and S.~Thakur, \emph{{A Charged
  Membrane Paradigm at Large D}},
  \href{http://dx.doi.org/10.1007/JHEP04(2016)128}{\emph{JHEP} {\bf 04} (2016)
  128}, [\href{http://arxiv.org/abs/1511.03432}{{\tt 1511.03432}}].

\bibitem{Dandekar:2016fvw}
Y.~Dandekar, A.~De, S.~Mazumdar, S.~Minwalla and A.~Saha, \emph{{The large D
  black hole Membrane Paradigm at first subleading order}},
  \href{http://arxiv.org/abs/1607.06475}{{\tt 1607.06475}}.

\bibitem{Emparan:2015hwa}
R.~Emparan, T.~Shiromizu, R.~Suzuki, K.~Tanabe and T.~Tanaka, \emph{{Effective
  theory of Black Holes in the 1/D expansion}},
  \href{http://dx.doi.org/10.1007/JHEP06(2015)159}{\emph{JHEP} {\bf 06} (2015)
  159}, [\href{http://arxiv.org/abs/1504.06489}{{\tt 1504.06489}}].

\bibitem{Suzuki:2015iha}
R.~Suzuki and K.~Tanabe, \emph{{Stationary black holes: Large $D$ analysis}},
  \href{http://arxiv.org/abs/1505.01282}{{\tt 1505.01282}}.

\bibitem{Tanabe:2015hda}
K.~Tanabe, \emph{{Black rings at large D}},
  \href{http://arxiv.org/abs/1510.02200}{{\tt 1510.02200}}.

\bibitem{Emparan:2015gva}
R.~Emparan, R.~Suzuki and K.~Tanabe, \emph{{Evolution and endpoint of the black
  string instability: Large D solution}},
  \href{http://dx.doi.org/10.1103/PhysRevLett.115.091102}{\emph{Phys. Rev.
  Lett.} {\bf 115} (2015) 091102}, [\href{http://arxiv.org/abs/1506.06772}{{\tt
  1506.06772}}].

\bibitem{Emparan:2016sjk}
R.~Emparan, K.~Izumi, R.~Luna, R.~Suzuki and K.~Tanabe, \emph{{Hydro-elastic
  Complementarity in Black Branes at large D}},
  \href{http://dx.doi.org/10.1007/JHEP06(2016)117}{\emph{JHEP} {\bf 06} (2016)
  117}, [\href{http://arxiv.org/abs/1602.05752}{{\tt 1602.05752}}].

\bibitem{Suzuki:2015axa}
R.~Suzuki and K.~Tanabe, \emph{{Non-uniform black strings and the critical
  dimension in the $1/D$ expansion}},
  \href{http://dx.doi.org/10.1007/JHEP10(2015)107}{\emph{JHEP} {\bf 10} (2015)
  107}, [\href{http://arxiv.org/abs/1506.01890}{{\tt 1506.01890}}].

\bibitem{Tanabe:2016pjr}
K.~Tanabe, \emph{{Elastic instability of black rings at large D}},
  \href{http://arxiv.org/abs/1605.08116}{{\tt 1605.08116}}.

\bibitem{Tanabe:2016opw}
K.~Tanabe, \emph{{Charged rotating black holes at large D}},
  \href{http://arxiv.org/abs/1605.08854}{{\tt 1605.08854}}.

\bibitem{Sadhu:2016ynd}
A.~Sadhu and V.~Suneeta, \emph{{Nonspherically symmetric black string
  perturbations in the large dimension limit}},
  \href{http://dx.doi.org/10.1103/PhysRevD.93.124002}{\emph{Phys. Rev.} {\bf
  D93} (2016) 124002}, [\href{http://arxiv.org/abs/1604.00595}{{\tt
  1604.00595}}].

\bibitem{Herzog:2016hob}
C.~P. Herzog, M.~Spillane and A.~Yarom, \emph{{The holographic dual of a
  Riemann problem in a large number of dimensions}},
  \href{http://arxiv.org/abs/1605.01404}{{\tt 1605.01404}}.

\bibitem{Rozali:2016yhw}
M.~Rozali and A.~Vincart-Emard, \emph{{On Brane Instabilities in the Large $D$
  Limit}},  \href{http://arxiv.org/abs/1607.01747}{{\tt 1607.01747}}.

\bibitem{Chen:2015fuf}
B.~Chen, Z.-Y. Fan, P.~Li and W.~Ye, \emph{{Quasinormal modes of Gauss-Bonnet
  black holes at large D}},
  \href{http://dx.doi.org/10.1007/JHEP01(2016)085}{\emph{JHEP} {\bf 01} (2016)
  085}, [\href{http://arxiv.org/abs/1511.08706}{{\tt 1511.08706}}].

\bibitem{Chen:2016fuy}
B.~Chen and P.-C. Li, \emph{{Instability of Charged Gauss-Bonnet Black Hole in
  de Sitter Spacetime at Large $D$}},
  \href{http://arxiv.org/abs/1607.04713}{{\tt 1607.04713}}.

\bibitem{Bhattacharyya:2008kq}
S.~Bhattacharyya, S.~Minwalla and S.~R. Wadia, \emph{{The Incompressible
  Non-Relativistic Navier-Stokes Equation from Gravity}},
  \href{http://dx.doi.org/10.1088/1126-6708/2009/08/059}{\emph{JHEP} {\bf 08}
  (2009) 059}, [\href{http://arxiv.org/abs/0810.1545}{{\tt 0810.1545}}].

\bibitem{Camps:2010br}
J.~Camps, R.~Emparan and N.~Haddad, \emph{{Black Brane Viscosity and the
  Gregory-Laflamme Instability}},
  \href{http://dx.doi.org/10.1007/JHEP05(2010)042}{\emph{JHEP} {\bf 05} (2010)
  042}, [\href{http://arxiv.org/abs/1003.3636}{{\tt 1003.3636}}].

\bibitem{Caldarelli:2012hy}
M.~M. Caldarelli, J.~Camps, B.~Goutéraux and K.~Skenderis,
  \emph{{AdS/Ricci-flat correspondence and the Gregory-Laflamme instability}},
  \href{http://dx.doi.org/10.1103/PhysRevD.87.061502}{\emph{Phys. Rev.} {\bf
  D87} (2013) 061502}, [\href{http://arxiv.org/abs/1211.2815}{{\tt
  1211.2815}}].

\end{thebibliography}\endgroup
\end{document}